\documentclass[dvips,twocolumn]{revtex4}
\usepackage{amsmath,amssymb,amsthm}
\usepackage[pdftex]{graphicx}

\begin{document}

\title{Universal Quantum Computation with Continuous-Variable Abelian
Anyons}
\author{Darran F. Milne$^{1}$, Natalia V. Korolkova$^{1}$, Peter van
Loock$^{2}$}

\affiliation{$^{1}$School of Physics and Astronomy, University of St.
Andrews, North Haugh, St. Andrews, Fife, KY16 9SS, Scotland, \\
$^{2}$Optical Quantum Information Theory Group, Max-Planck Institute for the Science of Light, G\"{u}nther-Scharowsky-Str.1/Bau 26 and Institute of
Theoretical Physics, Universit\"{a}t Erlangen-N\"{u}rnberg, Staudtstr. 7/B2, 91058 Erlangen, Germany}

\begin{abstract}

We describe how continuous-variable abelian anyons, created on the
surface of a continuous-variable analogue of Kitaev's lattice model
can be utilized for quantum computation. In particular, we derive
protocols for the implementation of quantum gates using topological
operations. We find that the topological operations alone are
insufficient for universal quantum computation which leads us to study
additional non-topological operations such as offline squeezing and
single-mode measurements. It is shown that these in conjunction with a
non-Gaussian element allow for universal quantum computation using
continuous-variable abelian anyons.

\end{abstract}
\maketitle

\section{Introduction}

The efficient storage and processing of quantum information is a major
open problem for quantum computation (QC). One of the main routes to
protect quantum information against random errors introduced via
environmental decoherence and noise is the use of quantum error correcting
(QEC) codes \cite{qec1,qec2,NandC,Shor}.
Essentially, such schemes attempt to correct errors with clever
`software' design. In contrast, recently, a new approach has been suggested in which
information is intrinsically protected by storing it as a non-local topological
degree of freedom. This approach is known as Topological Quantum Computation
(TQC). In these schemes, ideally, the protection is an intrinsic property of the
physical system, being inaccessible to noise. Moreover, any quantum
gates that are implemented by non-trivial topological operations are
also protected.

A recent development in TQC is the realisation of certain exotic states of matter that occur in two dimensions, the so-called anyon states, which use these topological properties of a system to provide a natural medium for storing and manipulating quantum information \cite{wilzcek,preskill}. Anyons are quasiparticle excitations that exhibit fractional statistics, i.e., when two anyons are exchanged the quantum state acquires a phase shift corresponding to the fractional spin of the anyonic states. These anyonic states first emerged in connection with the fractional quantum Hall effect \cite{fqhe}, which occurs in a two-dimensional electron gas at low temperatures. Anyons have been extensively studied for their fundamental interest, but recently it has been shown that anyons are a useful resource for fault-tolerant topological quantum computation \cite{Kitaev,nonabelian1,nonabelian2}. Anyons fall into two general classes; abelian and non-abelian. Non-abelian anyons have received the most attention as it has been shown that some species, notably the Fibonacci anyons \cite{nonabelian1}, are a resource for universal QC over discrete variables using just their braiding and fusion operations alone. However, it has been shown that the operations available to abelian anyons can also provide a universal gate set \cite{wootton, jiang, bond}, but certain operations such as single-qubit rotations, must be carried out using non-topological methods.

A simple model from which one can produce
abelian anyons was proposed by Kitaev for spin $1/2$ systems
\cite{Kitaev,Kitaev2}. However, this surface code model involves
suitable combinations of four-body interactions which are difficult to
achieve experimentally. In \cite{qubitmodel} it was shown that this
code can be created efficiently from a two-dimensional cluster state
by selectively measuring out single spins. This protocol was extended
to the Continuous-Variable (CV) regime in \cite{CVanyon} where it was
shown that from a CV cluster state, a CV analogue of the Kitaev model
can be constructed. It was demonstrated that the continuous excitations
above this ground state are CV abelian anyons with non-trivial
braiding statistics.
Here we study the computational power of these CV anyons. We show how
to create gates based on their topological properties and find that we
can implement single-mode phase-space displacements as well as two-mode
controlled phase-space displacements, where both the control and the
target can be either in the computational or in the conjugate basis.
However, analogous to the qubit case, we find that the topological
operations for the CV abelian anyons do not form a sufficient gate set
for universal QC and so we include certain non-topological operations
to supplement the topological operations in order to complete the gate set.

We begin, in Sec.~II, with a short introduction to quantum computation
over continuous variables. In Sec.~III, we review briefly how to
construct the CV Kitaev ground state from a CV cluster state and
discuss the fusion and braiding properties of the anyonic excitations.
In Sec.~IV we examine the range of Clifford group operations that are
achievable with the CV anyons by topological and non-topological
actions. The Clifford gates alone are not enough for universal CV QC
and hence in Sec.~V, we study the effect on the anyons of applying a cubic
phase gate to the Kitaev ground state. In Secs.~VI and VII, we
investigate the effect of finite squeezing of the resource state on
the excitations and computational model. We conclude in Sec.~VIII.

\section{Continuous-variable quantum computation}

Quantum logic over continuous variables generalizes the qubit Pauli
$X$ and $Z$ operators to the Weyl-Heisenberg (WH) group \cite{CV}, the
group of phase space displacements. This is a Lie group with
generators $\hat{x}=(\hat{a}+\hat{a}^{\dagger})/\sqrt{2}$ and
$\hat{p}=i(\hat{a}^{\dagger}-\hat{a})/\sqrt{2}$ representing, for
instance, a single quantized mode (qumode) of the electromagnetic
field. These operators satisfy the canonical commutation relation
$[\hat{x},\hat{p}]=i$, equivalent to position and momentum. In what
follows, we refer to $\hat{x}$ and $\hat{p}$ as position and momentum.
The single-mode WH operators are defined as  $X(s)=e^{-is \hat{p}}$
and $Z(t)=e^{it \hat{x}}$, $s,t \in \mathbb{R}$. The WH operator
$X(s)$ is the position-translation operator, which acts on the
computational basis of position eigenstates $\{|x\rangle ;x\in
\mathbb{R} \}$ as $X(s)|x\rangle = |x+s\rangle$. $Z(t)$ is the
momentum-translation operator, which acts on the momentum eigenstates
$\{|p\rangle ; p \in \mathbb{R} \}$
as $Z(t)|p\rangle = |p+t\rangle$. These operators are non-commutative and obey the identity
\begin{equation}
X(s)Z(t)=e^{-ist}Z(t)X(s).
\end{equation}
In the following we will show how a CV computational model can be
constructed using CV abelian anyons.

\section{Generation of Anyonic states from CV graph states}

We review the protocol of \cite{CVanyon}, for the generation of the CV
Kitaev ground state. This scheme is based on the first Kitaev lattice
model which is a spin Hamiltonian for a two-dimensional square
lattice. In Kitaev's scheme, a qubit (for instance, a spin $1/2$
particle) is associated with each edge of a square lattice, for which the model
Hamiltonian is given by
\begin{equation}
H=-\sum_{s} A_s - \sum_f B_f,
\end{equation}
where $A_s=\prod_{j\in star(s)} X_j$, and $B_f=\prod_{j\in
\partial(f)}Z_j$. Here $\partial(f)$ denotes the boundary spins of a
plaquette and the operators $X$ and $Z$ are the standard Pauli matrix
operators $\sigma_x$ and $\sigma_z$.

\begin{figure}[htp]
\begin{center}
\includegraphics[width=7cm]{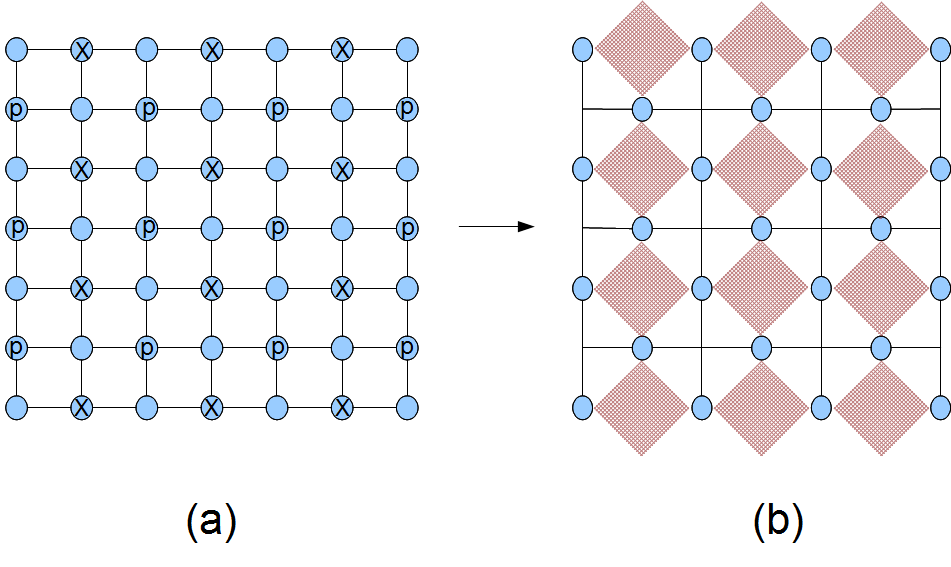}
\end{center}
\caption{(a) The measurement pattern to prepare the anyonic ground
state from a two-dimensional cluster state, where $X$ and $P$ denote a
single-mode measurement in the position and momentum bases,
respectively. (b) The CV surface code model.}
\label{Meas}
\end{figure}

These four-body interactions can be produced experimentally by cooling
the system to its ground state, but a more practical approach suggested in
\cite{qubitmodel} is to create the ground state dynamically from graph
or cluster states. In \cite{CVanyon} this model was generalized to CV
states which we turn to now.

In the ideal case, CV cluster states are prepared from a collection of
$N$ zero-momentum eigenstates \cite{cluster1,peter1,peter2,peter3}, which we write
as $|0\rangle_{p}^{\otimes N}$, where the $p$-subscripted kets satisfy
$\hat{p}|s\rangle_p = s|s\rangle_p$. These states are then entangled
via a collection of controlled-Z operations, denoted
$C_Z=\exp(ig\hat{x}_i\otimes \hat{x}_j)$, where $g \in \mathbb{R}$ is
the strength of the interactions (we will assume $g=1$ throughout).
Labelling the nodes of the graph in some arbitrary order, we can
define a symmetric \textit{adjacency matrix} $\textbf{A}=\textbf{A}^T$
whose $(j,k)th$ entry $A_{jk}$ is equal to the weight of the edge
linking node $j$ to node $k$ (with no edge corresponding to a weight
$0$). Note in the ideal case the diagonal entries are all zero since
we do not allow for self loops in the graph \cite{graphcalc}. The
collection of controlled-Z operations used to make the CV cluster
state is then a function of \textbf{A}, denoted
$C_Z[\textbf{A}]$. The CV cluster state with the graph
\textbf{A} is then
\begin{eqnarray}
|\psi_A \rangle &=& C_Z[\textbf{A}]|0\rangle_p^{\otimes N}
\nonumber \\
&=& \prod_{j,k=1}^{N} \exp \left(\frac{i}{2} A_{jk}\hat{x}_j \hat{x}_k
\right)|0\rangle_p^{\otimes N} \nonumber \\
&=& \exp \left(\frac{i}{2} \hat{\textbf{x}}^T \textbf{A}
\hat{\textbf{x}}\right)|0\rangle_p^{\otimes N},
\end{eqnarray}
where $\hat{\textbf{x}}=(\hat{x}_1,...,\hat{x}_N)^T$ is a column
vector of position operators.
Ideal CV cluster states in the unphysical limit of infinite squeezing
satisfy a set of \textit{nullifier} relations,
which can be written as
\begin{equation}
(\hat{\textbf{p}}-\textbf{A}\hat{\textbf{x}})|\psi_A \rangle = 0,
\end{equation}
where $\hat{\textbf{p}}=(\hat{p}_1,...,\hat{p}_N)^T$ is a column
vector of momentum operators. This represents $N$ independent
equations, one for each component of the vector $(\hat{\textbf{p}}-
\textbf{A} \hat{\textbf{x}})$, which are the nullifiers for $|\psi_A
\rangle$, because that state is a simultaneous zero-eigenstate of them.
The nullifiers are written explicitly as
\begin{equation}
\hat{g}_a=(\hat{p}_a - \sum_{b\in N_a} \hat{x}_b) \rightarrow 0, \;\;
\forall a \in G,
\end{equation}
where the modes $a \in G$ correspond to the vertices of the graph of
$N$ modes and the modes $b \in N_a$ are the nearest neighbours of mode
$a$.  An efficient way of representing CV cluster states is via the
stabilizer formalism. A state $|\phi\rangle$ is stabilized by an
operator $K$ if it is an eigenstate of $K$ with unit eigenvalue. If
such a set exists for a given state, then we call the state a
stabilizer state, and we may use the generators of its stabilizer
group to uniquely specify it. The stabilizer operator for a
cluster state is $G_a(\xi)=\exp(-i\xi \hat{g}_a)=X_a(\xi)\prod_{b \in
N_a} Z_b(\xi)$.

To construct the CV Kitaev code state from a CV cluster state, we make
single-mode measurements on every second mode in either the position
or momentum basis, with a subsequent Fourier transform, $F$, for all
the remaining modes which leaves us with a new graph state
$|\psi\rangle$ (Fig. 1). This new state is described by the set of
correlations of the form $\{\hat{a}_s =(
\hat{p}_{s,1}+\hat{p}_{s,2}+\hat{p}_{s,3}+\hat{p}_{s,4} ) \rightarrow
0,\; \hat{b}_{f}=(\hat{x}_{f,1}-\hat{x}_{f,2} +\hat{x}_{f,3} -
\hat{x}_{f,4}) \rightarrow 0\}$ where $s$ and $f$ label star and
plaquettes, respectively, and the indices $1,..,4$ of the position and
momentum operators denote those modes located at a common star or at
the boundary of a common plaquette. The new stabilizer operators that
describe this state, $\{A_{s}(\xi)=\exp(-i\xi \hat{a}_s)=\prod_{j\in
star(s)}X_{s,j}(\xi),\; B_f(\eta)=\exp(-i\eta \hat{b}_f)=\prod_{j\in
\partial(f)}Z_{f,j}((-1)^j \eta)\}$ with $\xi, \eta \in \mathbb{R}$,
are analogous to the first Kitaev model for a two-dimensional spin
lattice. Since these new stabilizer operators commute, the new ground
state corresponds to an anyonic ground state with
$A_s(\xi)|\psi\rangle = |\psi\rangle$ and
$B_f(\xi)|\psi\rangle=|\psi\rangle$ for all stars $s$ and plaquettes
$f$, in the limit of infinite squeezing.

We regard continuous excitations above the pre-prepared ground state
$|\psi\rangle$ as CV anyons, produced by applying $Z$ and $X$
operators on the ground state. Specifically, the position-translation
operator applied to some mode of the lattice creates a pair of
$m$-type anyons on adjacent plaquettes, i.e., $|m((-1)^d t)\rangle =
X(t)|\psi\rangle$ ($d \in \{1,2\}$), where $d=1$ means the relevant mode lies
on the vertical edges, and $d=2$ refers to the horizontal edges). An
$e$-type pair of anyons is created on adjacent vertices of the lattice
by $|e(s)\rangle = Z(s)|\psi\rangle$. We define a computational basis
$|r\rangle_{v/f}$ on vertices, $v$, or plaquettes, $f$, with $r\in \mathbb{R}$, composed of a pair of anyons
as
\begin{align}
&|r\rangle_{v} = |e(r)\rangle_{v_1}|e(-r)\rangle_{v_2},\nonumber \\
&|r\rangle_{f} = |m(r)\rangle_{f_1}|m(-r)\rangle_{f_2}.
\end{align}
The fusion rules $e(s) \times e(t) = e(s+t), \;\; m(s) \times m(t) =
m(s+t), \;\; e(0)\times e(s) = e(s), \;\; m(0) \times m(s)=m(s)$
describe the outcome of combining two anyons. By application of a
sequence of $X$ and $Z$ operators we can braid the anyons. For
example, consider an initial state $|\psi_{ini}\rangle =
Z_i(s)|\psi\rangle= |e(s)\rangle$. If an anyon of type $m$ is at a
neighbouring plaquette, it can be moved around $e$ along a path
generated by successive application of $X(t)$ on the four modes of the
star. The final state is
\begin{eqnarray}\label{braid}
|\psi_{fin}\rangle &=& X_1(t) X_2(t) X_3(t) X_4(t)|\psi_{ini}\rangle
\nonumber\\
                   &=& e^{-ist}Z_i(s)[X_1(t) X_2(t) X_3(t)
X_4(t)|\psi\rangle] \nonumber \\
                   &=& e^{-ist}|\psi_{ini}\rangle.
\end{eqnarray}
The phase factor is known as the topological phase factor, which
reveals the presence of enclosed anyons. Note that the same phase
factor would be acquired by the state independent of the path the
braiding anyon follows. This topological character reveals the
potential robustness of operations with CV anyons and their use as a
resource for fault-tolerant quantum computation.

\section{Clifford Gates}

We now examine the computational power of the CV abelian anyons. In
particular we show how the Clifford group operations are achieved
using both topological and non-topological means.
The set $\{Z(s), F,P(\eta),C_Z;s,\eta\in \mathbb{R} \}$ generates the
Clifford group \cite{CV}, where $P(\eta)= \exp[i(\eta/2)\hat{x}^2]$ is a CV
squeezing gate, $F=\exp[i\pi/4 (\hat{x}^2 + \hat{p}^2)]$ is the Fourier Transform
operator, and $C_Z=\exp(ig\hat{x}_i\otimes \hat{x}_j)$ is the
controlled-Z gate, as defined before. Transformations within the Clifford group correspond to Gaussian transformations mapping Gaussian states
onto Gaussian states. We see below that the topological operations
available to CV abelian anyons are not sufficient to generate the
entire group and we will require non-topological operations to complete the
set.

\subsection{Topological Operations}

Our first topological operations are quadrature displacements, $Z(s)=e^{is\hat{x}}$. Phase-space displacements are achieved through the creation and fusion of anyons. It is easy to see that creation of an anyon results in a displacement away from the ground state. For non-trivial displacements we fuse anyons of the same type, created by displacements on modes $i$ and $j$. The anyon on site $j$ can then be moved to site $i$ to implement fusion,
\begin{eqnarray}
|e(s)\rangle \times |e(t)\rangle &=& Z_i(s)|\psi\rangle \times Z_j(t)|\psi\rangle \nonumber \\
                                 &=& [Z(s) \times Z(t)]_i|\psi\rangle \nonumber \\
                                 &=& e^{i(s+t)\hat{x_i}}|\psi\rangle.
\end{eqnarray}
To act our quadrature displacement on the computational basis we must ensure that both anyons in the produced pair are fused with its counterpart. Hence the effect of our displacement on the computational basis is
\begin{equation}
|r+s\rangle_{v} = |e(r + s)\rangle_{v_1}|e(-(r+s))\rangle_{v_2}.
\end{equation}
The change in the computational basis for the $m$-type anyons follows similarly. We can extend this to the two mode SUM gate (Fig. 2), which is a
controlled displacement $C_X = e^{-i\hat{x}_i \otimes \hat{p}_j}$, i.e., $|x\rangle_1 |y\rangle_2 \rightarrow |x\rangle_1 |x + y\rangle_2$. We affect the SUM gate by fusing one of the anyons from the first mode with an anyon from the second mode. This results in a displacement of the second mode dependant on the state of the first:
\begin{eqnarray}
|s\rangle_1 |t\rangle_2 &=& (|e(s)\rangle |e(-s)\rangle)_1 (|e(-
t)\rangle |e(t)\rangle)_2 \nonumber \\
&\rightarrow&  |e(-s)\rangle_1 |e(s+t)\rangle_2 \nonumber \\
&=& |-s\rangle_1 |s+t\rangle_2,
\end{eqnarray}
where we treat the second anyon of mode two as a spectator anyon that
will be annihilated at the end of the computation and the control is
left in the original state (up to a sign change).

\begin{figure}[htp]
\begin{center}
\includegraphics[width=7cm]{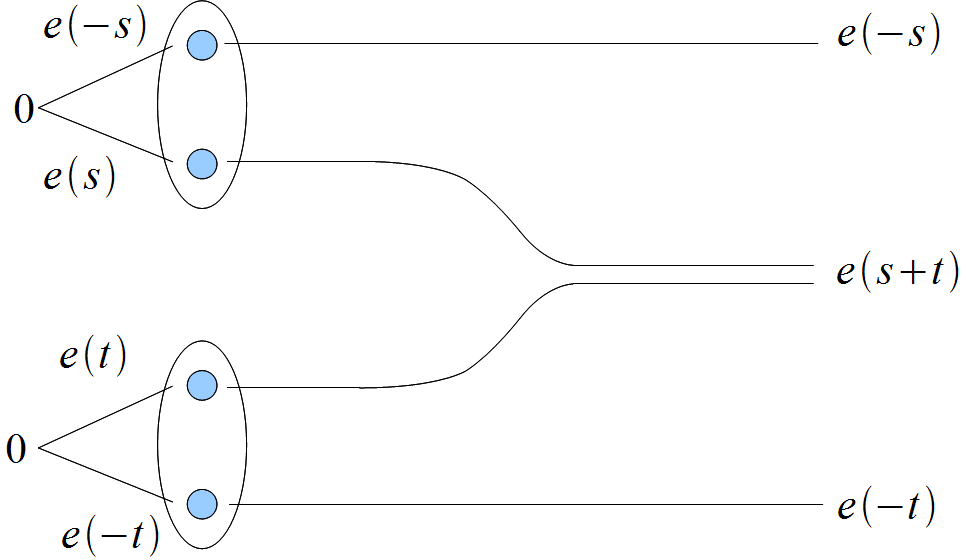}
\end{center}
\caption{The two-mode SUM gate. An anyon from the top pair is fused to
produce a controlled displacement on the lower pair.}
\label{SUM}
\end{figure}

We can also perform a controlled-Z gate (Fig. 3), which acts on the
computational basis as $|x\rangle_1 |y\rangle_2 \rightarrow
|x\rangle_1 e^{i\phi}|y\rangle_2$. This time we affect this
transformation by braiding the anyons. For example to perform a
controlled displacement in the conjugate basis on an $m$-type anyon we
braid an $e$-type anyon around it. From equation (\ref{braid}), we see
that the state picks up a phase dependant on the anyonic states.

\begin{figure}[htp]
\begin{center}
\includegraphics[width=7cm]{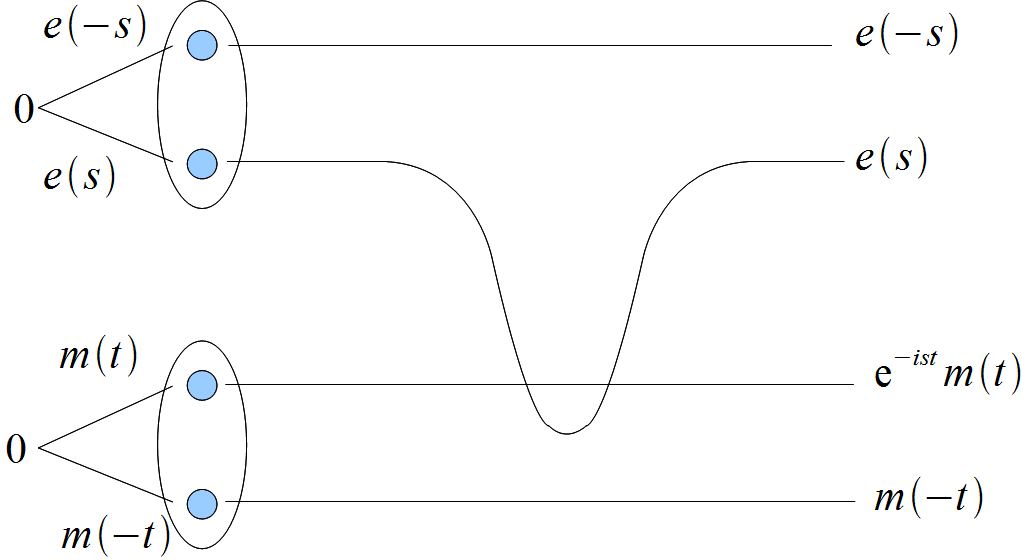}
\end{center}
\caption{A controlled phase gate $C_Z$. Braiding an anyon from the top pair
with the bottom pair produces a phase change dependant on the state of
the state of the anyons.}
\label{Phase}
\end{figure}
Hence, we have found that the controlled shift operation $C_X$,
$e^{i\hat{x}_i\hat{p}_j}$, corresponds to partial fusions of $e$-type
anyon pairs, while the controlled shift operation $C_Z$,
$e^{i\hat{x}_i\hat{x}_j}$, corresponds to partial braiding between $e$-
and $m$-type anyon pairs. Similarly, we note that the operations
$e^{i\hat{p}_i\hat{p}_j}$ and $e^{i\hat{p}_i\hat{x}_j}$ are achieved
by partial braiding between $m$ and $e$-type anyon pairs and partial
fusion of $m$-anyon pairs, respectively.

\subsection{Non-Topological Operations}

Above we saw how we can implement single mode displacements,
displacements in the conjugate basis and two-mode controlled displacement gates. Now we go beyond simple first-moment Gaussian operations and consider the manipulation of second moments. In particular, we seek to complete the set of Clifford gates
by the inclusion of a squeezer and a Fourier transform. The squeezing
operation compresses the position quadrature by a factor $\eta$
while stretching the conjugate quadrature by $1/\eta$. We cannot
directly squeeze our anyons since the only action we can take on the
anyons is fusion and braiding. Instead, we squeeze on some mode $i$ of the ground state,
\begin{equation}
|\tilde{\psi}\rangle = P_i(\eta)|\psi\rangle.
\end{equation}
Then creating an anyon on this squeezed ground state mode and commuting
through gives us
\begin{eqnarray}
Z_i(s)|\tilde{\psi}\rangle &=& Z_i(s)P_i(\eta)|\psi\rangle \nonumber\\
                         &=& P_i(\eta)e^{i2\eta s \hat{x}_i}Z_i(s)|\psi\rangle.
\end{eqnarray}
We find that squeezing the ground state is equivalent to squeezing the
anyon with an additional phase space displacement. This squeezing
operation combined with a measurement in the $X$ basis can be used to
implement a Fourier transform $F$. The action of $F$ is to switch
between the position and momentum bases, i.e., $F|x\rangle_x =
|x\rangle_p$. This corresponds to a generalization of the Hadamard
gate for qubits. To perform $F$ on our anyons, we begin by preparing a
zero-momentum squeezed ground state $|0\rangle_p$. We proved above
that up to a displacement, any squeezer on the ground state acts on the
anyonic excitations in the same way. Hence we can produce zero-momentum anyons. For example, consider an $m$-type anyon in the
computational basis $|\psi' \rangle = X(s) \int dt f(t) |t\rangle_q$. We fuse this
anyon with the momentum squeezed anyon:
\begin{eqnarray}
\text{SUM}[|\psi'\rangle \times |0\rangle_p]&=& \text{SUM}[X(s)|\psi\rangle
\times |0\rangle_p] \nonumber \\
                                            &=& \text{SUM} \left[\int dt f(t)
X(s)|t\rangle_q |0\rangle_p \right] \nonumber\\  &=& \int dt f(t)
|t+s\rangle_q |t+s\rangle_p.
\end{eqnarray}
Then, performing a measurement of $\hat{p}$ with outcome $m$ on the
mode corresponding to the first anyon collapses this to
\begin{eqnarray}
\int dt f(t)e^{i(t+s)m} |s+t\rangle_p = X(m)F|\psi'\rangle.
\end{eqnarray}
We see that the effect of this procedure is to apply a Fourier
transform modulo a known quadrature displacement. This completes our
set of Clifford gates, and so with appropriate non-topological
operations we can apply any Gaussian transformation. As stated above,
Gaussian transformations are not sufficient for universal QC and we
address the question of universality in the next section.

\section{Non-Clifford gates and universality}

For our abelian anyon computational scheme, we require either a non-Gaussian gate or a non-Gaussian resource. In a similar fashion to our
squeezing operation, we apply a transformation known as a cubic phase \cite{gkp}, $V(\gamma) = e^{i\gamma \hat{x}^3}$, $\gamma \in \mathbb{R}$, to some mode $i$ of the ground state. Then we commute $V(\gamma)$ through the $X$ and $Z$ operators to find the effect on the anyons:
\begin{equation}
|\psi_{c}\rangle = V_i(\gamma)|\psi\rangle.
\end{equation}
Now applying $X$ and $Z$ operators to create $e$- and $m$-type anyons
and commuting the cubic phase gate through we find
\begin{align}
X_i(s)Z_i(t)|\psi_{c}\rangle &= X_i(s)Z_i(t)V_i(\gamma)|\psi\rangle \nonumber \\ &=V_i(\gamma)e^{-i(s\hat{p}_i - t\hat{x}_i + 3\gamma s \hat{x}_i^2)} X_i(s)Z_i(t)|\psi \rangle.
\end{align}
Hence we find that applying a cubic phase to the ground state is
equivalent to applying the cubic phase to the anyons with extra
displacements and squeezing operations. This produces non-Gaussian
anyons that can then be used to obtain a universal gate set. To
conclude our discussion, we attempt to provide a classification of our
gate operations. We found that robust topological operations
correspond to controlled or uncontrolled WH gates. Non-robust,
non-topological Clifford operations correspond to symplectic operations.
Non-robust, non-topological, non-Clifford operations correspond to
non-WH, non-symplectic operations. In group theoretical terms, all
operations that are not elements of the normalizer of the WH group
(Clifford group) cannot be topologically realized using abelian
anyons. Of those operations that are elements of the normalizer
of WH, only the elements of the normal subgroup of the Clifford group
can be topologically realized using abelian anyons; those Clifford
elements that are not elements of the normal subgroup cannot be
realized topologically using abelian anyons
\footnote[1]{Note that in standard CV encoding, the $C_x$ and $C_z$ gates are
symplectic operations involving squeezers and beam splitters.
In the abelian-anyon encoding, however, they would only result in controlled shifts.}.

\section{Physical States - Finite Squeezing}

So far, we have only considered the generation of anyonic statistics
on an infinitely squeezed ground state. However, this state is highly
unphysical. Here we extend our model to include finite squeezing of
the initial state and show what effects this has on our computational
model.

A method to extend the graph representations from ideal (infinitely
squeezed) CV cluster states to their finitely squeezed Gaussian
approximations was given in \cite{graphcalc}. There it was shown that
the nullifier formalism for CV cluster states can be extended to
general Gaussian pure states using the simple replacement of the CV
cluster state graph \textbf{A} with the Gaussian graph \textbf{Z}, so
that $(\hat{\textbf{p}} - \textbf{Z}\hat{\textbf{q}})|\phi_Z\rangle =
0$ with the new non-Hermitian nullifiers defined as
\begin{equation}
g_k = \hat{p}_k - ie^{-2r_k}\hat{x}_k - \sum_{l \in N_k}\hat{x}_l,
\;\;\;\; \forall k.
\end{equation}
Then the adjacency matrix \textbf{Z} for a Gaussian pure state is a
complex matrix with imaginary diagonal entries, $ie^{-2r_k}$,
corresponding to self-loops on the modes, and the
remaining entries either $0$ or $1$ depending on the particular CV
cluster state.
Starting from an $N$-mode square cluster state defined by this
nullifier and carrying out the measurement pattern described for the
ideal cluster, we generate the finitely squeezed Kitaev lattice with
complex nullifiers $\hat{a}'_s=\hat{a}_s=
(\hat{p}_{s,1}+\hat{p}_{s,2}+\hat{p}_{s,3}+\hat{p}_{s,4})= 0$ and
$\hat{b}'_f = \hat{b}_f -ie^{-2r_1}\hat{p}_{f,1}+
ie^{-2r_2}\hat{p}_{f,2}-ie^{-2r_3}\hat{p}_{f,3}+
ie^{-2r_4}\hat{p}_{f,4} = 0$,
for the remaining modes, where $s$ and $f$ label the stars and
plaquettes, respectively, and $\partial f$ denotes the boundary of a
face. Comparing with the nullifiers in the infinitely squeezed limit, we
observe that finite squeezing introduces extra imaginary terms to the
plaquette nullifiers.
The stabilizers corresponding to these complex nullifiers are
$A'_s(\xi)=e^{-i\xi \hat{a}'_s}=\prod_{j \in star(s)}
X_{s,j}(\xi)$ and $B'_f(\eta)=e^{-i\eta
\hat{b}'_f}=\prod_{j \in \partial (f)}Z_{f,j}((-1)^j \eta)\exp[\eta (-1)^{j} e^{-2r_j}(\eta +\hat{p}_j)]$. These operators correspond to the
Kitaev model with an extra complex term, but note that these reduce
exactly to the Kitaev stabilizers as the squeezing parameter $r
\rightarrow \infty$.
These new stabilizers still commute and so the new state $|\phi \rangle$ corresponds to the anyonic ground state with $A'_s(\xi)|\phi
\rangle = |\phi \rangle$ and $B'_f(\eta)|\phi \rangle = |\phi \rangle$. In order to see the effects of finite squeezing, we apply the unphysical stabilizers to the physical ground state,
\begin{align}
B_{f}&(\eta)|\phi \rangle = \exp[-i \eta \hat{b}_f]|\phi \rangle \nonumber \\
=& \exp \left[\eta\sum_{j=1}^4 (-1)^{j+1} e^{-2r_j}(\hat{p}_j+\eta)\right] \nonumber \\
\times & \exp \left[-\eta \sum_{j=1}^4 (-1)^{j+1} e^{-2r_j}(\hat{p}_j +\eta)\right] \exp[-i \eta \hat{b}_f]|\phi \rangle \nonumber \\
=& \exp \left[-\eta \sum_{j=1}^4 (-1)^{j+1} e^{-2r_j}(\hat{p}_j+\eta) \right]B'_f(\eta)|\phi \rangle \nonumber \\
=& \exp \left[-\eta \sum_{j=1}^{4} (-1)^{j+1} e^{-2r_j}(\hat{p}_j+\eta)\right]|\phi \rangle,
\end{align}
where we dropped the subscripts $f$ of the momentum operators for simplicity.
We observe that finite squeezing of the ground state violates the unphysical stabilizer conditions by an imaginary phase, $\sim i \eta^2$, and by imaginary position shifts, $\sim i\eta$. Having derived the form of our finitely squeezed CV lattice, we now turn to anyonic excitations and basic braiding operations.

\section{Anyonic creation and braiding on finitely squeezed lattice}

By applying single-mode operations to the ground state, we can examine the excitations above the physical ground state. The effects of finite squeezing on the creation of anyonic excitations are revealed when we calculate the violation of the finitely squeezed nullifiers due to the application of $X_i(t)$ and $Z_i(t)$ on some mode of the physical ground state defined by
\begin{equation}
A'_s(\xi)|\phi \rangle = e^{-i \xi \hat{a}'_s}|\phi \rangle= |\phi\rangle,
\end{equation}
\begin{equation}
B'_f(\eta)|\phi \rangle = e^{-i \eta \hat{b}'_f}|\phi \rangle = |\phi\rangle.
\end{equation}
Note that now, due to the non-Hermiticity of the physical nullifiers,
their violations through anyonic excitations may in general be complex.
An excitation of the vertex ground state due to $Z_j(t)$ corresponds to
\begin{equation}\label{finsqanyonbegin}
A'_s(\xi)[Z_j(t)|\phi\rangle] = e^{-i \xi \hat{a}'_s}e^{it \hat{x}_j}|\phi\rangle,
\end{equation}
but $\hat{a}'_s = \hat{a}_s$, as the finite squeezing has no effect on the vertex nullifier. Hence,
\begin{align}
 A'_s(\xi)[Z_j(t)|\phi\rangle]&= e^{-i \xi \hat{a}_s}e^{it \hat{x}_j}|\phi\rangle \nonumber \\
 &= e^{\xi t[\hat{a}_s,\hat{x}_j]}Z_j(t)A_s(\xi)|\phi\rangle \nonumber \\
 &= e^{-i \xi t}Z_j(t)A'_s(\xi)|\phi\rangle \nonumber \\
 &= e^{-i\xi t}[Z_j(t)|\phi \rangle], \forall\xi\in\mathbb{R}.
\end{align}
Just as in the infinitely squeezed case, this yields a stabilizer violation of $t$. The $Z_j(t)$ applied to the plaquette stabilizers gives
\begin{align}
B'_f(\eta)[Z_j(t)|\phi\rangle]  &= e^{-i \eta \hat{b}'_f}e^{it\hat{x}_j}|\phi \rangle \nonumber \\
																&= e^{\eta t[\hat{b}'_f,\hat{x}_j]}Z_j(t)B'_f(\eta)|\phi\rangle \nonumber \\
																&= e^{\eta t[-ie^{-2r_j}\hat{p}_{f,j},\hat{x}_j]}[Z_j(t)|\phi \rangle] \nonumber \\
																&= e^{- \eta t e^{-2r_j}}[Z_j(t)|\phi \rangle], \forall\eta\in\mathbb{R}.
\end{align}
This differs from the infinitely squeezed case (where we had no violation at all), with an imaginary nullifier violation of $it e^{-2r_j}$. This time physical anyons may appear as complex violations of the ground-state stabilizers. Applying $X_j(t)$ to the ground state yields,
\begin{align}
A'_s(\xi)[X_j(t)|\phi\rangle]	&= e^{-i\xi \hat{a}'_s}e^{-it\hat{p}_j}|\phi \rangle \nonumber \\
														&= e^{-i \xi \hat{a}_s}e^{-it\hat{p}_j}|\phi\rangle \nonumber \\
														&= X_j(t)A'_s(\xi)|\phi\rangle \nonumber \\
														&= [X_j(t)|\phi\rangle], \forall\xi\in\mathbb{R},
\end{align}
so no stabilizer violation occurs. Finally, the plaquette stabilizer gives
\begin{align}
B'_f(\eta)[X_j(t)|\phi\rangle]  &= e^{-i\eta \hat{b}'_f}e^{-it\hat{p}_j}|\phi\rangle \nonumber \\
															&= e^{-\eta t [\hat{b}'_f,\hat{p}_j]}X_j(t)B'_f(\eta)|\phi\rangle \nonumber \\
															&= e^{-\eta t[\hat{b}_f,\hat{p}_j]}X_j(t)B'_f(\eta)|\phi\rangle \nonumber \\
															&= e^{\pm i \eta t}[X_j(t)|\phi\rangle], \forall\eta\in\mathbb{R},
\end{align}
which corresponds to a violation of $\pm t$.

In summary, finite squeezing gives us a violation of the plaquette ground state stabilizer due to the action of $Z(t)$, whereas it did not exhibit violations in the infinitely squeezed case. The vertex stabilizers are unaffected and yield violations of the same form as in the infinitely squeezed case. This suggests that any topological operations carried out on $e$ anyons are topologically protected from error. However, the
only operation possible on star anyons alone is the SUM gate which is not on its own sufficient for quantum computation. The anyons are now represented in general by complex violations. These may then said to be new types of excitations produced through finite squeezing which we call \textit{complex anyons}.

The effects on the braiding procedure and hence gate operations are determined by generating vertex and plaquette anyons and guiding them around each other in closed loops.
\begin{align}
|\phi_{fin} \rangle &= [Z_1(-t)Z_2(t)Z_3(-t)Z_4(t)]|\phi_{ini} \rangle \nonumber \\
										&= [Z_1(-t)Z_2(t)Z_3(-t)Z_4(t)]X_k(s)|\phi \rangle .
\end{align}
Commuting through to enact the braid yields
\begin{align}
|\phi_{fin}\rangle =& \exp[ist]X_k(s)[Z_1(-t)Z_2(t)Z_3(-t)Z_4(t)] |\phi \rangle \nonumber
\end{align}
\begin{align}
=	\exp[ist]X_k(s)\exp \left[-t \sum_{j=1}^4 (-1)^{j+1}e^{-2r_j} (\hat{p}_j+t)\right] B'_f(t)|\phi\rangle.
\end{align}
Using our definition of the ground state, $B'_f(t)|\phi \rangle = |\phi\rangle$, we obtain
\begin{align}
|\phi_{fin}\rangle  =& \exp[ist]X_k(s) \exp \left[-t \sum_j (-1)^{j+1}e^{-2r_j}(\hat{p}_j +t)\right] |\phi\rangle \nonumber \\
										=& \exp[ist] \exp \left[-t \sum_j (-1)^{j+1}e^{-2r_j}(\hat{p}_j+t)\right]|\phi_{ini}\rangle.
\end{align}
We can express the term proportional to $t$ as an imaginary displacement,
\begin{align}									
|\phi_{fin}\rangle =& \exp[ist] \exp \left[-t^2 \sum_j (-1)^{j+1}e^{-2r_j}\right] \nonumber \\
									  & \times\prod_j X_j \left(-it(-1)^{j+1}e^{-2r_j} \right)|\phi_{ini}\rangle.
\end{align}
As in the infinitely squeezed case, we observe a phase change of  $e^{ist}$, but this time, for finite squeezing,
we have an extra imaginary displacement and a term proportional to $t^2$ (corresponding to an extra imaginary
phase, similar to what we had before for the ground-state plaquette stabilizers with finite squeezing).
We may call the combination of these terms the \textit{topological factor for the braiding
of finitely squeezed anyons}, and we note that this factor would not be obtained if the initial states were unexcited. We can absorb the complex displacement into the definition of the ground state such that the nullifer $\hat{b}_f$ is no longer zero, but has an imaginary violation $\hat{b}_f = it\sum_j (-1)^{j+1}e^{-2r_j}$. Then $\hat{b}_f = s+ it\sum_j (-1)^{j+1}e^{-2r_j} = s'$ is the nullifier corresponding to a finitely squeezed $m$-anyon. The topological phase produced when braided with an $e$-type anyon is then
\begin{equation}\label{finsqanyonend}
\exp[is't]=\exp[i(s+it\sum_j (-1)^{j+1}e^{-2r_j})t].
\end{equation}
This is one of the central results of this paper, extending the simple (infinite squeezing) factor $e^{ist}$ of \cite{CVanyon} to the realistic case of finite squeezing. Similar to the infinitely squeezed case, the state can acquire any phase, but now the phase is modified by extra factors due to finite squeezing.

This should not affect topological gate operations since these displacements and dampings are taken into account through the definition of the ground state and they depend on the known squeezing $r_j$ available at each qumode. Hence we have shown that topological operations on CV abelian anyons are protected from errors due to finite squeezing of the initial ground state. Note that this result goes beyond that of
Ref.~\cite{CVanyon}. There it was argued that in the case of finite squeezing, excited (anyonic) states can be experimentally  distinguished from the ground state and, similarly, the effects of braiding loops can still be detected, provided the corresponding phase-space displacements are sufficiently large. In our treatment,
such a requirement is unnecessary. Using the complex-nullifier formalism, we find that any finite-squeezing effects on the 1st-moment-shifts can be absorbed into the definition of the excited states as well as into the topological phases.
However, note that these re-defined WH frames would always depend on the explicit values of the corresponding
anyonic excitations and anyon-braidings [i.e., the $t$-dependencies in Eqs.~(\ref{finsqanyonbegin}-\ref{finsqanyonend})].

As a result, all those gates shown to be implementable in a topological fashion turn out to be robust against finite squeezing errors. Nonetheless, non-topological
operations such as the Fourier transform do pick up extra errors due to finite squeezing since they rely on the ability to create zero-momentum eigenstates. In fact, all these non-topological gates include 2nd-or higher-moment manipulations which will be affected by the finite squeezing of the initial graph states. Similarly, of course, the entanglement of the ground and excited states, becoming manifest through nonclassical 2nd-moment correlations, does depend on the finite squeezing of the initial states.

\section{conclusion}

We have shown that abelian anyons generated on an ideal CV Kitaev
lattice are a useful resource for continuous-variable quantum
computation. We have described the quantum gates that can be achieved
by topological operations alone and found this did not form a
sufficiently powerful gate set. By including offline squeezing,
measurements, and cubic phase gates, we have shown that the
computational power can be increased significantly. These additional
resources provide a universal gate set for the CV anyons.
It has therefore been possible for us to give a classification of the topological operations available to CV abelian anyons. We have shown that braiding and fusion only account for controlled and uncontrolled WH gates, non-topological operations are essential to complete the Clifford group, and a further non-Gaussian element is required to achieve universality. We note that the topological controlled displacements only give entanglement
(including 2nd-moment non-classical correlations) if we have access to superpositions of anyonic excitations, e.g. in a bi-anyonic vacuum state, $\int ds\; dt \; e^{-s^2 -t^2}|s\rangle |t\rangle$. Such an extension will give rise to a more general model including non-abelian anyons.

We have also considered the effect of finite squeezing on the CV
Kitaev lattice and found that the excitations on this physical ground state
correspond to Gaussian anyon states. These physical anyons, mathematically
represented by complex stabilizer violations,
can perform fault-tolerant topological operations that are protected against errors due
to finite squeezing. However, we cannot create a fully protected gate
set, since some gates require non-topological operations. Nonetheless, we have been
able to identify those operations which are protected,
and in a future work, we shall attempt to present a universal, topological gate set using non-abelian anyons.
\\

This research has been supported by the EU STREP project COMPAS FP7-
ICT-2007-C-212008 under the FET-Open Programme,
by the Scottish Universities Physics Alliance (SUPA) and by the
Engineering and Physical Sciences Research Council (EPSRC). DFM
acknowledges the financial support from the Deutscher Akademischer
Austausch Dienst (DAAD). PvL acknowledges support from the Emmy Noether program of the DFG.

\end{document}